\documentclass[12pt]{article}
\usepackage{amssymb}
\usepackage{amsfonts}
\usepackage{amsmath}
\usepackage[mathscr]{eucal}
\usepackage{amssymb}
\usepackage{amsthm}
\usepackage{bbold}
\usepackage{bm}
\usepackage{graphicx}
\usepackage{caption}

\numberwithin{equation}{section}
\newcommand{\be}{\begin{equation}}
\newcommand{\ee}{\end{equation}}
\newcommand{\bea}{\begin{eqnarray}}
\newcommand{\eea}{\end{eqnarray}}

\usepackage{MnSymbol}

\newcounter{rown}

\begin{document}
\begin{titlepage}
\vspace*{0.1cm}

\begin{center}
{\LARGE\bf Model of massless relativistic particle with}

\vspace{0.2cm}

{\LARGE\bf continuous spin and its twistorial description}

\vspace{1.2cm}

{\large\bf I.L. Buchbinder$^{1,2,3,4}$,\,\, S. Fedoruk$^4$,\,\,  A.P. Isaev$^{4,5}$,\,\, A. Rusnak$^6$}

\vskip 1cm

\ $^1${\it Department of Theoretical Physics,
Tomsk State Pedagogical University, \\
Tomsk 634041, Russia}, \\
{\tt joseph@tspu.edu.ru}

\vskip 0.4cm

\ $^2${\it National Research Tomsk State University,\\
Tomsk 634050, Russia}\\

\vskip 0.4cm

\ $^3${\it Departamento de F\'isica, ICE, Universidade Federal de Juiz
de Fora,\\
Campus Universit\'ario-Juiz de Fora, 36036-900, MG, Brazil}\\

\vskip 0.4cm

\ $^4${\it Bogoliubov Laboratory of Theoretical Physics,
Joint Institute for Nuclear Research, \\
Joliot-Curie 6, 141980 Dubna, Moscow Region, Russia}, \\
{\tt fedoruk@theor.jinr.ru, isaevap@theor.jinr.ru}

\vskip 0.4cm

\ $^5${\it St.\,Petersburg Department of the Steklov Mathematical Institute \\ of Russian
Academy of Sciences, \\
Fontanka 27, 191023 St. Petersburg, Russia}

\vskip 0.4cm

\ $^6${\it Department of Physics {\rm \&} Technology,
Karazin Kharkov National University, \\
Svobody Sq. 4, UA 61022 Kharkov, Ukraine}\\
{\tt rusnak.aa@phystech.edu}

\end{center}

\vspace{0.6cm}

\nopagebreak

\begin{abstract}
\noindent We propose a new world-line Lagrangian model of the $D{=}\,4$
massless relativistic particle with continuous spin and develop its
twistorial formulation. The description uses two Penrose twistors
subjected to four first class constraints. After the first quantization of
the world-line twistorial model, the wave function is defined by an
unconstrained function on the two-dimensional complex affine plane. We
find the twistor transform that determines the space-time field of
the continuous spin particle through the corresponding twistor one,
which plays the role of a prepotential. It is shown that this
space-time field is an exact solution of the space-time constraints
defining the irreducible massless representation of the Poincar\'{e}
group with continuous spin.
\end{abstract}

\vspace{0.8cm}


\phantom{Keywords: }

\newpage

\end{titlepage}
\setcounter{footnote}{0}
\setcounter{equation}{0}

\section{Introduction}

\quad\, Among the unitary irreducible representations of the
Poincar\'{e} group \cite{Wigner39,Wigner47,BargWigner}, the massless
representations with continuous spin look somewhat
unusual from a physical point of view. In contrast to other
physical states having a positive or zero mass, they include an
infinite number of massless spin states (see, for example,
\cite{Iv-Mack}). This property of the continuous spin particles is
very attractive at the present time by reason of the intensive
development of higher-spin theory \cite{Vas1989,Vas1991,Vas1992}
(see also the reviews \cite{Vas2001,Vas2002,Vas2005}) and
explanation of its relation to the string theory (see the recent paper
\cite{Vas2018} and references in it). In addition to this important
circumstance, massless states of continuous spin have a description
in terms of string-localized fields \cite{MundSY}  and allow for
extensions to the supermultiplets \cite{BKRX}. Therefore, lately a
lot of research has been carried out on the  continuous spin particles
(see, for example,
\cite{BekBoul,BekMou,SchToro13a,SchToro13b,SchToro13c,SchToro15,Riv,Mets16,Mets17,BekSk,HabZin,AlkGr}
and references therein). These studies allow us to take a fresh look
at the problems in the theory of continuous spin particles, taking
into account the achievements of the string theory and the higher
spin field theory.

Up to now, all the considered formulations of the continuous spin
particles have been the space-time ones. That is, the description of
the corresponding massless states was carried out by using
coordinate and momentum vectors, as well as some additional
variables. The authors of most papers on this question use the
formulation with additional vector variables, which was proposed by
Wigner and Bargmann \cite{Wigner47,BargWigner}. The fields of the
continuous spin in this formulation are described by the function
$\Phi(x,y)$ defined on the space which is parametrized by two
commuting four-vectors $x^m$ and $y^m$.\footnote{The space-time
metric is $\eta_{mn}={\rm diag}(+1,-1,-1,-1)$. The totally antisymmetric
tensor $\epsilon_{mnkl}$ has the component $\epsilon_{0123}=-1$.
The two-component Weyl spinor indices are raised and lowered by
$\epsilon_{\alpha\beta}$, $\epsilon^{\alpha\beta}$,
$\epsilon_{\dot\alpha\dot\beta}$, $\epsilon^{\dot\alpha\dot\beta}$
with the nonvanishing components
$\epsilon_{12}=-\epsilon_{21}=\epsilon^{21}=-\epsilon^{12}=1$:
$\psi_\alpha=\epsilon_{\alpha\beta}\psi^\beta$,
$\psi^\alpha=\epsilon^{\alpha\beta}\psi_\beta$, etc. Relativistic
$\sigma$-matrices are $ (\sigma_m)_{\alpha\dot\beta}=({\bf
1_2};\sigma_1,\sigma_2,\sigma_3)_{\alpha\dot\beta} $, where
$\sigma_1,\sigma_2,\sigma_3$ are the Pauli matrices. The matrices $
(\tilde\sigma_{m})^{\dot\alpha\beta}=\epsilon^{\dot\alpha\dot\delta}\epsilon^{\beta\gamma}
(\sigma_m)_{\gamma\dot\delta}=({\bf
1_2};-\sigma_1,-\sigma_2,-\sigma_3)^{\dot\alpha\beta} $ satisfy $
\sigma^m_{\alpha\dot\gamma}\tilde\sigma^{n\,\dot\gamma\beta}+\sigma^m_{\alpha\dot\gamma}\tilde\sigma^{n\,\dot\gamma\beta}
=2\,\eta^{mn}\delta^\beta_\alpha$ and
$\sigma^m_{\alpha\dot\beta}\tilde\sigma_n^{\dot\beta\alpha}=2\,\delta^m_n$.
The link between Minkowski four-vectors and spinorial quantities is
given by
$A_{\alpha\dot\beta}={\textstyle\frac{1}{\sqrt{2}}}\,A_m(\sigma^m)_{\alpha\dot\beta}$,
$A^{\dot\alpha\beta}={\textstyle\frac{1}{\sqrt{2}}}\,A_m(\tilde\sigma^m)^{\dot\alpha\beta}$,
$A_m
={\textstyle\frac{1}{\sqrt{2}}}\,A_{\alpha\dot\beta}(\tilde\sigma_m)^{\dot\beta\alpha}$,
so that $ A^m B_m =A_{\alpha\dot\beta}B^{\dot\beta\alpha}$. We use
also the following quantities:
$A^{\dot\beta}_{\alpha}=\epsilon^{\dot\beta\dot\gamma}A_{\alpha\dot\gamma}=\epsilon_{\alpha\gamma}A^{\dot\beta\gamma}$,
$A_{\dot\beta}^{\alpha}=\epsilon_{\dot\beta\dot\gamma}A^{\dot\gamma\alpha}=\epsilon^{\alpha\gamma}A_{\gamma\dot\beta}$
(see the details, e.g., in \cite{BK}).} One of these vectors, the
vector $x^m$, defines the position coordinates on Minkowski space,
whereas additional vector variables $y^m$ are intended to describe
the spin degrees of freedom of massless particles. The conditions
describing the irreducible representation of the Poincar\'{e} group with
continuous spin can be written in terms of the function $\Phi(x,y)$ in
the form \cite{Wigner47,BargWigner}\footnote{We note that these
conditions were given in the papers \cite{Wigner47,BargWigner} in a
somewhat different but equivalent form.}
\begin{eqnarray}
\frac{\partial}{\partial x^m}\frac{\partial}{\partial x_m} \ \Phi &=& 0 \,, \label{eq-sp}\\
\frac{\partial}{\partial x^m}\frac{\partial}{\partial y_m} \ \Phi &=& 0 \,,   \label{eq-sp-1}\\
\frac{\partial}{\partial y^m}\frac{\partial}{\partial y_m} \ \Phi &=& \mu^2\,  \Phi\,,  \label{eq-sp-2}\\
-\,i\,y^m\,\frac{\partial}{\partial x^m} \ \Phi &=& \Phi \,,  \label{eq-sp-3}
\end{eqnarray}
where the constant $\mu\in\mathbb{R}$ is a dimensionful
parameter defining the value  of the four-order Casimir operator for
the massless representations of the Poinc\'{a}re group. In
\cite{Iv-Mack}, in addition to the coordinate and energy-momentum
vectors, additional spinor variables were used. However,  massless states
of finite spin (helicity) have an alternative description in terms of
twistor variables \cite{Pen67,PC72,PenRin}. Moreover, the twistor
formalism makes it possible to obtain a transparent treatment of
many aspects of space-time description, such as, for example, the
description of symmetries in the higher spin theory
\cite{BL99,BLS00,V+,FI}. In addition, the twistor approach is used
not only for massless states of finite spin. For example, there are
twistor formulations of massive particles with fixed spin (see,
for example, \cite{FZ03,FFLM06,FL14,IsP} and references therein) as
well as massive particles of higher spins \cite{AFIL}.

In this paper, we construct a twistorial formulation of a massless
particle of continuous spin in the usual four-dimensional space-time and
find its link with the field formulation. This will give additional
understanding of this system and also restore its further relations
with other theories of relativistic particles.

The plan of this paper is as follows. In the next section, we
formulate the Lagrangian dynamical system which reproduces after
quantization the space-time description \eqref{eq-sp}-\eqref{eq-sp-3} of the continuous spin
fields. The formulation under consideration uses in the Hamiltonian
approach an additional pair of four-vector variables. In Sect.\,3, we
construct a twistorial formulation, which is equivalent to the space-time
one of the previous section. For twistorial resolution of all space-time
momenta we introduce two twistors. We present also the incidence
relations that give the relationship of the space-time phase space
coordinates with the twistor variables. As a result, we establish a
full set of twistor constraints that describe the continuous spin
particle. In Sect.\,4, we make canonical quantization of the
twistorial model of the continuous spin particle. We find a twistorial
wave function (twistorial field)  of  the continuous spin particle.
This field is defined by an arbitrary function defined on the
two-dimensional complex affine plane. In Sect.\,5, we consider the
field twistor transform which reproduces space-time field  of the
continuous spin by the corresponding twistorial field. This space-time
field $\Phi(x,y)$ is the exact solution of equations
\eqref{eq-sp}-\eqref{eq-sp-3} by construction. In the last section, we
give some comments about the obtained results and present outlooks.

\setcounter{equation}{0}

\section{Space-time formulation}

\quad\, We begin with the construction of the new Lagrangian system for
the relativistic particle corresponding to the Wigner-Bargmann
space-time formulation of the irreducible continuous spin massless
representation. In this approach equations
\eqref{eq-sp}-\eqref{eq-sp-3} will arise as result of quantization
of the particle model.

Consider the one-dimensional dynamical model with the following
Lagrangian in the first-order formalism
\begin{equation}
\label{L-sp} L_{sp.-time}=p_m \dot x^m + w_m \dot y^m + e p_m p^m +
e_1 p_m w^m + e_2 \left(w_m w^m +\mu^2\right) + e_3 \left(p_m y^m
-1\right)
\end{equation}
and study the properties of this model. Here the quantities $p_m$
and $w_m$ are the momenta for the four-vector coordinates $x^m$ and
auxiliary vector variables $y^m$; their nonvanishing Poisson brackets are
\begin{equation}
\label{PB-sp} \left\{ x^m, p_n \right\}=\delta^m_n\,,\qquad
\left\{y^m, w_n \right\}=\delta^m_n\,.
\end{equation}
The parameter $\mu\in\mathbb{R}$ in the Lagrangian \eqref{L-sp} is a
constant and the $\tau$ is an evolution parameter: $x^m=x^m(\tau)$,
$p_m=p_m(\tau)$, etc. We use the standard notation for
$\tau$-derivatives: $\dot x^m:=dx^m/d\tau$, $\dot y^m:=dy^m/d\tau$.

The variables $e(\tau)$, $e_1(\tau)$, $e_2(\tau)$, $e_3(\tau)$ in the Lagrangian  \eqref{L-sp} are the Lagrange multipliers for the constraints
\cite{Wigner47,BargWigner} \footnote{
In \cite{Wigner47,BargWigner}, there were used dimensionless vectors $u^m=\mu y^m$, $\xi^m=\mu^{-1}w^m$,
in terms of which the constraints \eqref{const-sp-2}, \eqref{const-sp-3} take the form
$\xi_m \xi^m +1  \approx  0$, $p_m u^m -\mu  \approx  0$.}
\begin{eqnarray}
T &:=& p_m p^m \ \approx \ 0 \,, \label{const-sp}\\
T_1 &:=& p_m w^m \ \approx \ 0  \,,  \label{const-sp-1}\\
T_2 &:=& w_m w^m +\mu^2 \ \approx \ 0  \,,  \label{const-sp-2}\\
T_3 &:=& p_m y^m -1 \ \approx \ 0 \,.  \label{const-sp-3}
\end{eqnarray}
Nonvanishing Poisson brackets of the constraints \eqref{const-sp}-\eqref{const-sp-3} are
\begin{equation}
\label{nonv-PB-constr-sp}
\left\{ T_1, T_3 \right\}=-T\,,\qquad \left\{T_2, T_3\right\}=-2T_1\,.
\end{equation}
Therefore, all constraints \eqref{const-sp}-\eqref{const-sp-3} are first class.

The action ${\displaystyle S_{sp.-time}=\int d\tau L_{sp.-time}}$ with Lagrangian \eqref{L-sp} is invariant under
the transformations
\begin{equation}
\label{trans-sp}
\begin{array}{lll}
& \delta x^m = a_m+\omega_{mn} x^n\,,& \quad\delta p_m = \omega_{mn} p^n \,, \\
[6pt] & \delta y^m = \omega_{mn} y^n\,,& \quad\delta w_m =
\omega_{mn} w^n \,,
\end{array}
\end{equation}
where $a_m,\, \omega_{mn}=-\omega_{nm}\in\mathbb{R}$ are the
parameters of the global inhomogeneous Lorentz transformations. The
Noether charges for the transformations \eqref{trans-sp} take the
form
\begin{equation}
\label{Not-sp}
\begin{array}{rcl}
P_m &=& p_m \,, \\  [6pt]
M_{mn} &=& x_m p_n - x_n p_m + y_m w_n - y_n w_m
\end{array}
\end{equation}
and generate the Poincar\'{e} algebra with respect to Poisson brackets.

Due to the constraint \eqref{const-sp} the square of four-translations is zero,
\begin{equation}
\label{P2-sp}
P_m P^m\approx 0\,.
\end{equation}
Therefore, the model \eqref{L-sp} describes massless particle(s).

Due to the constraints \eqref{const-sp}-\eqref{const-sp-3} the square of the Pauli-Lubanski pseudovector
\begin{equation}
\label{W-def}
W_m=\frac12\,\varepsilon_{mnkl}P^n M^{kl}
\end{equation}
is equal to
\begin{equation}
\label{W2-sp}
W_m W^m=\frac12\, M_{nk}M^{nk}P_m P^m-M_{mk}M^{nl}P^k P_l\approx
-\mu^2\,.
\end{equation}
As result of this and \eqref{P2-sp}, we see that the model \eqref{L-sp}
indeed describes the massless particle with continuous spin. It is
noted that the vectors $w_m$ and $W_m=\varepsilon_{mnkl}p^n y^{l}w^{l}$
are not equal to each other, although both vectors are space-like
and have the equal lengths $w_m w^m=W_m W^m=-\mu^2$ on mass-shell.

After canonical quantization the constraints \eqref{const-sp}-\eqref{const-sp-3} yield
equations for the continuous spin fields \eqref{eq-sp}-\eqref{eq-sp-3}.
Additional coordinates $y^m$ in the arguments of these fields play the role of spin variables.

We note that the constraints \eqref{const-sp}-\eqref{const-sp-3} are also suitable for the description of
continuous spin particles in the Minkowski space-time of arbitrary dimension $D{=}\,d{+}1$
(see also the review~\cite{BekSk}).
In this paper, we consider a continuous spin  particle only in the four-dimensional space-time ($D{=}\,3{+}1$),
since the twistor formulation analyzed here is mostly well developed in the four-dimensional case.

At the end of this Section, we find a number of unconstrained
(physical) degrees of freedom in the model  \eqref{L-sp}. In this
model, we have four first class constraints
\eqref{const-sp}-\eqref{const-sp-3} and four gauge fixing
conditions. They eliminate eight degrees of freedom out of sixteen
phase variables $x^m$, $p_m$; $y^m$, $w_m$. Therefore, the model
\eqref{L-sp} is described by
\begin{equation}
n_{phys}=8
\label {phys-sp}
\end{equation}
phase variables. We will check the fulfillment of the condition
\eqref{phys-sp} in constructing a new twistorial formulation of the
continuous spin particle.

\setcounter{equation}{0}

\section{Twistorial formulation of continuous spin particle}

\quad\, We turn to constructing the  twistorial formulation of the system  \eqref{L-sp}.
To obtain this, we will follow the standard prescriptions of the twistor approach \cite{Pen67,PC72,PenRin}.
We stress that in the twistor formulation the models of the  massless particles describe the sector with energy of a definite sign.

The first step in this twistor program is to resolve the constraint \eqref{const-sp}.
Introducing commuting Weyl spinor $\pi_{\alpha}$, $\bar\pi_{\dot\alpha}=(\pi_{\alpha})^*$,
we represent light-like vector $p_m=p_{\alpha\dot\alpha}$ by the Cartan-Penrose relation \cite{Pen67,PC72,PenRin}
\begin{equation}
\label{p-resol}
p_{\alpha\dot\alpha}=\pi_{\alpha}\bar\pi_{\dot\alpha}\,.
\end{equation}
As a result of \eqref{p-resol} and $\pi^{\alpha}\pi_{\alpha}\equiv 0$, the constraint \eqref{const-sp} is satisfied automatically.
In addition, relation \eqref{p-resol} implies positive energy $p_0>0$ if the condition
$\pi_\alpha\neq (0,0)$ holds.
In the twistor formulation, the spinor $\pi_{\alpha}$ determines half of the components of the Penrose twistor.

The second step for obtaining the twistor formulation consists in a spinor representation of the four-vector components $w_m=w_{\,\alpha\dot\alpha}$
which should resolve equations \eqref{const-sp-1}, \eqref{const-sp-2}.
For this, it is necessary to introduce in addition one more commuting Weyl spinor $\rho_{\alpha}$, $\bar\rho_{\dot\alpha}=(\rho_{\alpha})^*$.
Then we represent $w_{\alpha\dot\alpha}$ in the form
\begin{equation}
\label{w-resol}
w_{\alpha\dot\alpha}=\pi_{\alpha}\bar\rho_{\dot\alpha}+\rho_{\alpha}\bar\pi_{\dot\alpha}\,.
\end{equation}
Expressions  \eqref{p-resol} and  \eqref{w-resol} solve the constraint \eqref{const-sp-1}.
Moreover, equality  \eqref{w-resol} leads to the following twistor expression for the square of the space-like vector $w_m$:
\begin{equation}
\label{w2-resol}
w_{\alpha\dot\alpha}w^{\alpha\dot\alpha}=-2\,\pi^{\alpha}\rho_{\alpha}\,\bar\rho_{\dot\alpha}\bar\pi^{\dot\alpha}=
-2|\pi^{\alpha}\rho_{\alpha}|^2\,.
\end{equation}
As a result, the constraint \eqref{const-sp-2} leads to the necessity of imposing in twistor space the following constraint:
\begin{equation}
\label{M-constr-def}
\mathcal{M}\ :=\ \pi^{\alpha}\rho_{\alpha}\,\bar\rho_{\dot\alpha}\bar\pi^{\dot\alpha}-M^2\ \approx\  0\,,
\end{equation}
where the real constant $M$ is defined by
\begin{equation}
\label{M-mu}
M^2=\mu^2/2\,.
\end{equation}
Without loss of generality, we will consider below the twistor system in which the constant $M$ is positive, $M>0$.

As the next step in our twistor program we introduce canonically conjugated momenta for
$\pi_{\alpha}$, $\bar\pi_{\dot\alpha}$ and $\rho_{\alpha}$, $\bar\rho_{\dot\alpha}$.
To do this, we represent the kinetic terms in  \eqref{L-sp}, $p_m \dot x^m + w_m \dot y^m$,
as the corresponding terms for twistorial spinors and their momenta.
We define spinorial momenta $\omega^{\alpha}$, $\bar \omega^{\dot\alpha}$ and $\eta^{\alpha}$, $\bar \eta^{\dot\alpha}$
by the following incidence relations:
\begin{equation}
\label{y-tw}
\omega^{\alpha}=\bar\pi_{\dot\alpha}x^{\dot\alpha\alpha}+\bar\rho_{\dot\alpha}y^{\dot\alpha\alpha}\,,\qquad
\bar \omega^{\dot\alpha}=x^{\dot\alpha\alpha}\pi_{\alpha}+y^{\dot\alpha\alpha}\rho_{\alpha}\,,
\end{equation}
\begin{equation}
\label{z-tw}
\eta^{\alpha}=\bar\pi_{\dot\alpha}y^{\dot\alpha\alpha}\,,\qquad
\bar \eta^{\dot\alpha}=y^{\dot\alpha\alpha}\pi_{\alpha}
\end{equation}
and find that, up to total derivative terms, they satisfy the relation
\begin{equation}
\label{kin-term-tw}
p_{\alpha\dot\alpha} \dot x^{\dot\alpha\alpha} + w_{\alpha\dot\alpha} \dot y^{\dot\alpha\alpha}=
\pi_{\alpha}\dot \omega^{\alpha} + \bar\pi_{\dot\alpha}\dot{\bar \omega}^{\dot\alpha}+
\rho_{\alpha}\dot \eta^{\alpha} + \bar\rho_{\dot\alpha}\dot{\bar \eta}^{\dot\alpha}\,.
\end{equation}
Thus, $(\omega^{\alpha}, \pi_{\alpha})$, $(\bar \omega^{\dot\alpha}, \bar\pi_{\dot\alpha})$,
$(\eta^{\alpha}, \rho_{\alpha})$, $(\bar \eta^{\dot\alpha}, \bar\rho_{\dot\alpha})$
are canonically conjugated pairs of the twistorial variables which obey the following Poisson brackets:
\begin{equation}
\label{tw-PB-sp}
\left\{ \omega^{\alpha}, \pi_{\beta} \right\}=\left\{ \eta^{\alpha}, \rho_{\beta} \right\}=\delta^\alpha_\beta\,,\qquad
\left\{\bar \omega^{\dot\alpha}, \bar\pi_{\dot\beta} \right\}=\left\{\bar \eta^{\dot\alpha}, \bar\rho_{\dot\beta} \right\}=\delta^{\dot\alpha}_{\dot\beta}\,.
\end{equation}

Using \eqref{p-resol} and \eqref{z-tw}, the last constraint \eqref{const-sp-3} in the space-time formulation
leads to the twistorial constraint
\begin{equation}
\label{const-tw-1}
\eta^{\alpha}\pi_{\alpha}+\bar\pi_{\dot\alpha}\bar \eta^{\dot\alpha}-2\approx 0\,.
\end{equation}
Furthermore, the incidence relations \eqref{z-tw} imply the twistor constraint
\begin{equation}
\label{const-tw-2}
\eta^{\alpha}\pi_{\alpha}-\bar\pi_{\dot\alpha}\bar \eta^{\dot\alpha}\approx 0\,.
\end{equation}
The sum and difference of \eqref{const-tw-1}, \eqref{const-tw-2} give equivalent constraints
\begin{equation}
\label{constr-tw}
{\mathcal{F}}\,:=\, \eta^{\alpha}\pi_{\alpha}-1\approx 0\,,\qquad
\bar{\mathcal{F}}\, :=\, \bar\pi_{\dot\alpha}\bar \eta^{\dot\alpha}-1\approx 0\,.
\end{equation}

Note that the incidence relations \eqref{y-tw}, \eqref{z-tw} imply also one more twistor constraint
\begin{equation}
\label{const-tw-4}
{\mathcal{U}}\, :=\, \omega^{\alpha}\pi_{\alpha}-\bar\pi_{\dot\alpha}\bar \omega^{\dot\alpha}
+\eta^{\alpha}\rho_{\alpha}-\bar\rho_{\dot\alpha}\bar \eta^{\dot\alpha}\approx 0\,.
\end{equation}

This finishes the twistorial formulation of the model  \eqref{L-sp}.
It is described by eight complex variables $\pi_{\alpha}$, $\omega^{\alpha}$, $\rho_{\alpha}$, $\eta^{\alpha}$,
which have with their complex conjugated variables
$\bar\pi_{\dot\alpha}$, $\bar \omega^{\dot\alpha}$; $\bar\rho_{\dot\alpha}$, $\bar \eta^{\dot\alpha}$
Poisson brackets \eqref{tw-PB-sp} and are subjected the constraints \eqref{M-constr-def}, \eqref{constr-tw} and \eqref{const-tw-4}.
The corresponding twistorial Lagrangian is
\begin{equation}
\label{L-tw}
L_{twistor}=
\pi_{\alpha}\dot \omega^{\alpha} \ + \ \bar\pi_{\dot\alpha}\dot{\bar \omega}^{\dot\alpha}  \ + \
\rho_{\alpha}\dot \eta^{\alpha}  \ + \  \bar\rho_{\dot\alpha}\dot{\bar \eta}^{\dot\alpha} \ + \
l\,{\mathcal{M}} \ + \  k\,{\mathcal{U}} \ + \  \ell\,{\mathcal{F}} \ +\  \bar\ell\,\bar{\mathcal{F}}\,,
\end{equation}
where $l(\tau)$, $k(\tau)$, $\ell(\tau)$, $\bar\ell(\tau)$ are the Lagrange multipliers
for the constraints \eqref{M-constr-def}, \eqref{constr-tw}, \eqref{const-tw-4}.

Four constraints \eqref{M-constr-def}, \eqref{constr-tw}, \eqref{const-tw-4} form an abelian algebra with respect to twistorial Poisson brackets  \eqref{tw-PB-sp}
and eliminate eight degrees of freedom from sixteen. As a result, the twistorial model  \eqref{L-tw} has the same
number of physical degrees of freedom  \eqref{phys-sp} as the model \eqref{L-sp} in the space-time formulation.

Let us check the preservation of the physical content of the theory when we switch to the twistor formulation \eqref{L-tw}.

Using (anti)self-dual spin-tensors
\begin{equation}
\label{Lor-asamdual}
M_{\alpha\beta} = \pi_{(\alpha}\omega_{\beta)}+\rho_{(\alpha}\eta_{\beta)} \,,\qquad
\bar M_{\dot\alpha\dot\beta}= \bar\pi_{(\dot\alpha}\bar\omega_{\dot\beta)}+\bar\rho_{(\dot\alpha}\bar\eta_{\dot\beta)}
\end{equation}
we represent the Noether charges of the Poincar\'{e} transformations \eqref{trans-sp} in the following form
\begin{equation}\label{M-tw}
\begin{array}{rcl}
M_{\alpha\dot\alpha\beta\dot\beta}&=&
\epsilon_{\dot\alpha\dot\beta} M_{\alpha\beta} + \epsilon_{\alpha\beta}\bar M_{\dot\alpha\dot\beta}  \,,\\ [6pt]
P_{\alpha\dot\alpha}&=& \pi_{\alpha} \bar\pi_{\dot\beta}  \,.
\end{array}
\end{equation}
These expressions can also be obtained by inserting expressions \eqref{p-resol}, \eqref{w-resol}, \eqref{y-tw}, \eqref{z-tw}
in charges \eqref{Not-sp}.
Then, in the Weyl-spinor notation, the Pauli-Lubanski vector \eqref{W-def} has the form
\begin{equation}
\label{PL-sp}
W_{\alpha\dot\alpha}=
-i\left( M_{\alpha\beta}P^{\beta}_{\dot\alpha}- \bar M_{\dot\alpha\dot\beta}P^{\dot\beta}_{\alpha}\right)
\end{equation}
and for the considered twistorial realization \eqref{Lor-asamdual}, \eqref{M-tw} we obtain the following expression:
\begin{equation}
\label{PL-tw}
W_{\alpha\dot\alpha}=
\Lambda P_{\alpha\dot\alpha}
-\frac{i}{2}\left[(\bar\pi_{\dot\beta}\bar\eta^{\dot\beta})\pi_{\alpha}\bar\rho_{\dot\alpha}
-(\pi_{\beta}\eta^{\beta})\rho_{\alpha}\bar\pi_{\dot\alpha} \right]
+\frac{i}{2}\left[(\bar\pi^{\dot\beta}\bar\rho_{\dot\beta})\pi_{\alpha}\bar\eta_{\dot\alpha}- (\pi^{\beta}\rho_{\beta})\eta_{\alpha}\bar\pi_{\dot\alpha}\right]\,,
\end{equation}
where
\begin{equation}
\label{hel-tw}
\Lambda:=\frac{i}{2}\left(\pi_{\beta}\omega^{\beta}- \bar\pi_{\dot\beta}\bar\omega^{\dot\beta}\right)\,.
\end{equation}
The square of the vector \eqref{PL-tw} is equal to
\begin{equation}
\label{sq-W}
W^2=W^{\alpha\dot\alpha}W_{\alpha\dot\alpha}=-2\,|\pi^{\alpha}\rho_{\alpha}|^2\,|\pi_{\beta}\eta^{\beta}|^2
\end{equation}
and, due to the constraints \eqref{M-constr-def}, \eqref{constr-tw}, we reproduce \eqref{W2-sp}: $W^2\approx -2M^2=-\mu^2$.
So the twistorial model \eqref{L-tw} describes massless particles of continuous spin.

Let us make some comments about our twistorial formulation of the continuous spin particle:
\begin{description}
\item[i)]
Introduce the matrix
\begin{equation}
\label{g-def}
\parallel\!\Pi_\alpha{}^{b}\!\parallel \ := \
\left(
\begin{array}{cc}
\pi_{1} & \rho_{1} \\
\pi_{2} & \rho_{2} \\
\end{array}
\right)
\,,
\end{equation}
where $b=1,2$, i.e.
\begin{equation}
\label{g-def1}
\left(\Pi_\alpha{}^{1},\Pi_\alpha{}^{2}\right) \ = \ \left(\pi_{\alpha},\rho_{\alpha}\right)
\,.
\end{equation}
We see that constraint \eqref{M-constr-def} is equivalent to
$\det \Pi=M\exp(i\varphi)$, where $\varphi$ defines the phase of the spinor contraction $\pi^{\alpha}\rho_{\alpha}$:
$\exp(2i\varphi)=(\pi^{\alpha}\rho_{\alpha})/(\bar\rho_{\dot\alpha}\bar\pi^{\dot\alpha})$.
So the matrix $M^{-1/2}\Pi_\alpha{}^{b}$
at $\varphi=0$ describes group manifold $\mathrm{SL}(2,\mathbb{C})$ parametrizing the spinor Lorentz harmonics \cite{Band,FedZim}.
\item[ii)]
The constraints \eqref{constr-tw} generate local transformations
\begin{equation}
\label{trans-loc}
\delta\rho_{\alpha}=n\pi_{\alpha}\,,\quad \delta\pi_{\alpha}=0\,,\quad
\delta\bar\rho_{\dot\alpha}=\bar n\bar\pi_{\dot\alpha}\,,\quad \delta\bar\pi_{\dot\alpha}=0\qquad
\Rightarrow\qquad \delta\Pi=\Pi\left(
                      \begin{array}{cc}
                        0 & n \\
                        0 & 0 \\
                      \end{array}
                    \right)
\,,
\end{equation}
where $n=n(\tau)$ is the local infinitesimal complex parameter. Thus, configuration space of the model is described
by the matrix \eqref{g-def} defined up to local transformations $g\to gN$, with the matrix
\begin{equation}
\label{N-def}
N=
\left(
\begin{array}{cc}
1 & n \\
0 & 1 \\
\end{array}
\right).
\end{equation}
These matrices for any $n\in\mathbb{C}$ form the Borel subgroup $B_+(2,\mathbb{C})$ of the Lorentz group $\mathrm{SL}(2,\mathbb{C})$.
The coset $\mathrm{SL}(2,\mathbb{C})/B_+(2,\mathbb{C})$ is the two-dimensional complex affine plane \cite{GelGV}.
This fact will be crucial in the analysis of the quantum spectrum of the  model \eqref{L-tw} considered in the next Section.
\item[iii)]
Spinors $\pi_{\alpha}$, $\bar\omega^{\dot\alpha}$ and $\rho_{\alpha}$, $\bar \eta^{\dot\alpha}$
form two Penrose twistors
\begin{equation}
\label{Z-tw-def}
Z_{A} := \left( \pi_\alpha, \bar{\omega}^{\dot\alpha} \right) , \qquad
Y_{A} := \left( \rho_\alpha, \bar{\eta}^{\dot\alpha} \right) .
\end{equation}
Conjugated spinors $\bar\pi_{\dot\alpha}$, $\omega^{\alpha}$ and $\bar\rho_{\dot\alpha}$, $\eta^{\alpha}$
are present in the definition of dual twistors
\begin{equation}
\label{Y-tw-def}
\bar Z^{A}:=
\left(\!
\begin{array}{c}
\omega^{\alpha} \\
-\bar{\pi}_{\dot{\alpha}} \\
\end{array}
\!\right);
\qquad
\bar Y^{A}:=
\left(\!
\begin{array}{c}
\eta^{\alpha} \\
-\bar{\rho}_{\dot{\alpha}} \\
\end{array}
\!\right),
\end{equation}
So in the description of the continuous spin particle it is necessary to have a bitwistor formulation
as opposed to the one-twistor description of a usual massless particle with fixed helicity.

Since norms of the twistors \eqref{Z-tw-def}, \eqref{Y-tw-def} are
\begin{equation}
\label{norm-tw}
\bar Z^{A}Z_A\, =\, \omega^{\alpha}\pi_{\alpha}-\bar\pi_{\dot\alpha}\bar \omega^{\dot\alpha}\,,\qquad
\bar Y^{A}Y_A\, =\, \eta^{\alpha}\rho_{\alpha}-\bar\rho_{\dot\alpha}\bar \eta^{\dot\alpha}\,,
\end{equation}
the constraint \eqref{const-tw-4} has the concise form
\begin{equation}
\label{const-tw-4-tw}
{\mathcal{U}}\, =\, \bar Z^{A}Z_A-\bar Y^{A}Y_A\,\approx\, 0\,.
\end{equation}
For a massless particle with fixed helicity, the norm of the twistor $Z$, which resolves particle momentum in terms of \eqref{p-resol},
is constant and defines helicity by the operator \eqref{hel-tw}:
$\Lambda=\frac{i}{2}\,\bar Z^{A}Z_A$. In our model, particle helicity is not fixed due to the constraint \eqref{const-tw-4-tw}.

\end{description}

\setcounter{equation}{0}
\section{Quantization of twistorial model and continuous spin twistor field}

\quad\, Before the quantization is performed,
we fix some gauges  and pass to an appropriate phase variable,
in which the constraints have a simple form.

Our twistor model of the continuous spin particle is described by the first class constraints
${\mathcal{M}}$, ${\mathcal{U}}$, ${\mathcal{F}}$, $\bar{\mathcal{F}}$
defined by expressions \eqref{M-constr-def}, \eqref{const-tw-4}, \eqref{constr-tw}.
Let us fix gauges for the constraints \eqref{M-constr-def} and \eqref{const-tw-4}.

The constraint \eqref{M-constr-def} involves only the norm $|\pi^{\alpha}\rho_{\alpha}|$.
Gauge fixing condition for this constraint must be canonically conjugated one.
The natural choice for such a constraint is the generator of conformal transformations of twistor components,
which has the form
\begin{equation}
\label{const-mass-fix}
{\mathcal{R}}\, :=\, \omega^{\alpha}\pi_{\alpha}+\bar\pi_{\dot\alpha}\bar \omega^{\dot\alpha}
+\eta^{\alpha}\rho_{\alpha}+\bar\rho_{\dot\alpha}\bar \eta^{\dot\alpha}\approx 0\,.
\end{equation}
We can check that Poisson bracket $\{{\mathcal{R}},{\mathcal{M}}\}=4{\mathcal{M}}+4M^2\approx4M^2\neq 0$ takes place, as required.

The constraint \eqref{const-tw-4} is the generator of phase transformations in twistor space.
In particular, these transformations act on a phase of the spinor contraction $\pi^{\alpha}\rho_{\alpha}$.
As result, we take the condition on this phase
\begin{equation}
\label{U-constr-def-fix}
\mathcal{K}\ :=\ \ln\left(\frac{\pi^{\alpha}\rho_{\alpha}}{\bar\rho_{\dot\alpha}\bar\pi^{\dot\alpha}}\right) \ \approx\  0
\end{equation}
as gauge fixing condition for the constraint \eqref{const-tw-4}.
Moreover, their Poisson bracket is $\{{\mathcal{U}},{\mathcal{K}}\}=4$.

The constraints \eqref{M-constr-def} and \eqref{U-constr-def-fix} are equivalent to the constraints
\begin{equation}
\label{N-constr-def-fix}
\mathcal{N}\ :=\ \pi^{\alpha}\rho_{\alpha}-M\ \approx\  0\,,\qquad
\bar{\mathcal{N}}\ :=\ \bar\rho_{\dot\alpha}\bar\pi^{\dot\alpha}-M\ \approx\  0\,.
\end{equation}
In addition, the sum and difference of the constraints \eqref{const-tw-4} and \eqref{const-mass-fix} are
\begin{equation}
\label{const-tw-fix}
{\mathcal{V}}\, :=\, \omega^{\alpha}\pi_{\alpha}
+\eta^{\alpha}\rho_{\alpha}\approx 0\,, \qquad
\bar{\mathcal{V}}\, :=\, \bar\pi_{\dot\alpha}\bar \omega^{\dot\alpha}
+\bar\rho_{\dot\alpha}\bar \eta^{\dot\alpha}\approx 0\,.
\end{equation}
Thus, the twistorial system of the continuous spin particle is described by the set of four second class constraints
\eqref{N-constr-def-fix}, \eqref{const-tw-fix} and two first class constraints \eqref{constr-tw}
\begin{equation}
\label{F-constr-def-fix}
{\mathcal{F}}\,:=\, \eta^{\alpha}\pi_{\alpha}-1\approx 0\,,\qquad
\bar{\mathcal{F}}\, :=\, \bar\pi_{\dot\alpha}\bar \eta^{\dot\alpha}-1\approx 0\,.
\end{equation}
Nonvanishing Poisson brackets of these constraints are
\begin{equation}
\label{PB-constr-new}
\{{\mathcal{V}},{\mathcal{N}}\}=2{\mathcal{N}}+2M\,,\qquad
\{\bar{\mathcal{V}},\bar{\mathcal{N}}\}=2\bar{\mathcal{N}}+2M\,.
\end{equation}

Let us pass to the variables in which the constraints \eqref{N-constr-def-fix}, \eqref{const-tw-fix}, \eqref{F-constr-def-fix} have a simple form.
Taking into account that the constraints \eqref{F-constr-def-fix} generate gauge transformations \eqref{trans-loc}
from parabolic subgroup of the $\mathrm{SL}(2,\mathbb{C})$ group,
we use the following expansion of the matrix \eqref{g-def}, \eqref{g-def1}:
\begin{equation}
\label{matrix-exp}
\parallel\!\Pi_\alpha{}^{b}\!\parallel \  = \
\sqrt{M}\left(
\begin{array}{cc}
z_{1} & 0 \\
z_{2} & s/z_{1} \\
\end{array}
\right)
\left(
\begin{array}{cc}
1 & t \\
0 & 1 \\
\end{array}
\right)
\,,
\end{equation}
where
\begin{equation}
\label{def-new-var-z}
z_{\alpha}= \pi_{\alpha}/\sqrt{M}\,,\qquad s=\pi^{\alpha}\rho_{\alpha}/M\,,\qquad
t=\rho_{1}/\pi_{1}
\end{equation}
are four dimensionless complex variables.
Relations \eqref{def-new-var-z} imply
\begin{equation}
\label{det-Pi}
\det\Pi=\pi^{\alpha}\rho_{\alpha}=Ms \,,\qquad
\det\bar\Pi=\bar\rho_{\dot\alpha}\bar\pi^{\dot\alpha}=M\bar s \,.
\end{equation}

We emphasize that the constraints \eqref{M-constr-def} imply
\begin{equation}
\label{lamda-no-0}
\left(\pi_{1},\pi_{2}\right)\neq \left(0,0\right)
\end{equation}
and therefore $\left(z_{1},z_{2}\right)\neq \left(0,0\right)$.
In the matrix expansion \eqref{matrix-exp}, we assume for definiteness that $z_{1} \neq 0$.

Let us make the canonical transformation from the phase spinor variables,
coordinates $\omega^{\alpha}$, $\eta^{\alpha}$, $\bar\omega^{\dot\alpha}$,  $\bar\eta^{\dot\alpha}$
and momenta $\pi_{\alpha}$, $\rho_{\alpha}$, $\bar\pi_{\dot\alpha}$,  $\bar\rho_{\dot\alpha}$,
to new phase space with coordinates $z_{\alpha}$, $s$, $t$, $\bar z_{\dot\alpha}$, $\bar s$, $\bar t$ and their momenta
$p_z^{\alpha}$, $p_s$, $p_t$, $\bar p_z^{\dot\alpha}$, $\bar p_s$, $\bar p_t$.
The main requirement for such a transformation is the fulfillment of relations \eqref{def-new-var-z} and their conjugate ones
\begin{equation}
\label{def-new-var-bz}
\bar z_{\dot\alpha}= \bar\pi_{\dot\alpha}/\sqrt{M}\,,\qquad \bar s=\bar\rho_{\dot\alpha}\bar\pi^{\dot\alpha}/M\,,\qquad
\bar t=\bar\rho_{1}/\bar\pi_{1}\,.
\end{equation}
The generating function of this transformation, depending on old and new momenta, has the following form:
\begin{equation}
\label{gen-func}
G(\pi_{\alpha},\bar\pi_{\dot\alpha},\rho_{\alpha},\bar\rho_{\dot\alpha};
p_z^{\alpha},\bar p_z^{\dot\alpha},p_s,\bar p_s,p_t,\bar p_t) \ = \ \frac{1}{\sqrt{M}}\,\pi_{\alpha}p_z^{\alpha} \ + \
\frac{\rho_1}{\pi_1}\,p_t \ + \ \frac{1}{M}\,\pi^{\alpha}\rho_{\alpha}p_s \ + \ \mbox{c.\,c.}
\end{equation}
The generating function \eqref{gen-func} leads to the relations
$z_\alpha=\partial G/\partial p_z^{\alpha}$, $\bar z_{\dot\alpha}=\partial G / \partial \bar p_z^{\dot\alpha}$,
$s=\partial G/\partial s$, $\bar s=\partial G / \partial \bar s$,
$t=\partial G/\partial t$, $\bar t=\partial G / \partial \bar t$.
In its turn, these relations allow us to obtain expressions \eqref{def-new-var-z} and \eqref{def-new-var-bz}.
Derivatives of the generating function \eqref{gen-func} with respect to old momenta give expressions for the old coordinates $\omega^{\alpha}$, $\eta^{\alpha}$, $\bar\omega^{\dot\alpha}$,  $\bar\eta^{\dot\alpha}$ through the remaining variables:
\begin{eqnarray}\label{mu-can}
\omega^\alpha &=& -\,\frac{\partial G}{\partial\pi_\alpha} \ = \ -\,\frac{1}{\sqrt{M}}\,p_z^{\alpha} \ + \
\frac{1}{M}\,\rho^{\alpha}p_s \ + \ \frac{\delta^{\alpha 1}\rho_1}{(\pi_1)^2}\,p_t  \,,\\
\eta^\alpha &=& -\,\frac{\partial G}{\partial\rho_\alpha} \ = \ -\,
\frac{1}{M}\,\pi^{\alpha}p_s \ - \ \frac{\delta^{\alpha 1}}{\pi_1}\,p_t  \,, \label{omega-can}
\end{eqnarray}
and complex conjugated relations.
From the definition of the canonical transformation with the generating function \eqref{gen-func} we
obtain that nonvanishing Poisson brackets of new variables are
\begin{equation}
\label{PB-n-var}
\left\{ z_{\alpha}, p_z^{\beta} \right\}=\delta_\alpha^\beta\,,\quad
\left\{\bar z_{\dot\alpha}, \bar p_z^{\dot\beta} \right\}=\delta_{\dot\alpha}^{\dot\beta}\,,\qquad
\left\{ s, p_s \right\}= \left\{\bar s, \bar p_s \right\}=1\,,\qquad
\left\{ t, p_t \right\}= \left\{\bar t, \bar p_t \right\}=1\,.
\end{equation}

In new variables the constraints  \eqref{N-constr-def-fix} take the form (see \eqref{det-Pi})
\begin{equation}
\label{N-const-def-fix}
s -1\approx 0\,,\qquad \bar s -1\approx 0\,.
\end{equation}
Moreover, equations \eqref{mu-can}, \eqref{omega-can} and c.c. lead to the following expressions
of the constraints  \eqref{const-tw-fix} in new variables:
\begin{equation}
\label{const-tw-fix-n}
-z_{\alpha}p_z^{\alpha} -
2sp_s\approx 0\,, \qquad
-\bar z_{\dot\alpha}\bar p_z^{\dot\alpha}
-2\bar s\bar p_s\approx 0\,.
\end{equation}
For definition of the constraints \eqref{F-constr-def-fix} in new variables we take into account the relations
$p_t=-\pi_{\alpha}\eta^{\alpha}$,
$p_t=-\bar\pi_{\dot\alpha}\bar\eta^{\dot\alpha}$,
which follow from  \eqref{omega-can} and c.c. Thus, in new variables the constraints \eqref{F-constr-def-fix} take the form
\begin{equation}
\label{constr-tw-new}
p_t+1\approx 0\,,\qquad
\bar p_t+1\approx 0\,.
\end{equation}
So in new variables all constraints (second class constraints  \eqref{N-const-def-fix}, \eqref{const-tw-fix-n} and
first class ones \eqref{constr-tw-new}) have a simple form.

We take account of the second class constraints  \eqref{N-const-def-fix}, \eqref{const-tw-fix-n} by introducing the Dirac bracket for them.
As a result, the second class constraints are now fulfilled in the strong sense
and remove the variables $s$, $\bar s$, $p_s$, $\bar p_s$ from the phase space of the model.
Due to the resolved form of the constraints \eqref{N-const-def-fix} (they depend only on variables $s$, $\bar s$),
the Dirac brackets for the rest of the phase space variables $z_{\alpha}$, $p_z^{\alpha}$, $t$, $p_t$ and c.c. coincide with their Poisson brackets.

We will perform quantization in the coordinate representation in which the realization of the operators
of phase variables is the following
(we take $\hbar=1$)
\begin{equation}
\label{realiz-var}
\begin{array}{l}
{\displaystyle \hat z_{\alpha}=z_{\alpha}\,, \quad \hat p_z^{\alpha} =-i\frac{\partial}{\partial z_{\alpha}}\,,\qquad
\hat {\bar z}_{\dot \alpha}=\bar z_{\dot \alpha}\,, \quad \hat {\bar p}_z^{\dot \alpha} =-i\frac{\partial}{\partial \bar z_{\dot \alpha}}
\,, } \\ [7pt]
{\displaystyle \hat t=t\,, \quad \hat p_t =-i\frac{\partial}{\partial t}\,,\qquad
\hat {\bar t} =\bar t\,, \quad \hat {\bar p}_t =-i\frac{\partial}{\partial \bar t}\,. }
\end{array}
\end{equation}
In this realization the wave function is
\begin{equation}
\label{wf-1}
\Psi=\Psi( z_{\alpha},\bar z_{\dot\alpha}, t, \bar t)\,.
\end{equation}

The wave function of physical states is subjected to the first class constraints \eqref{constr-tw-new} which lead to the equations
\begin{equation}
\label{quant-constr-new}
\frac{\partial}{\partial t}\,\Psi=\frac{\partial}{\partial \bar t}\Psi=-i\Psi\,.
\end{equation}
The solution of equations \eqref{quant-constr-new}, with taking into account the conditions of the second class constraints \eqref{N-const-def-fix}, is
\begin{equation}
\label{wf-sol}
\Psi \ = \ \delta(s-1)\,\delta(\bar s-1)\,e^{\displaystyle -i(t+\bar t)}\,\tilde\Psi( z_{\alpha},\bar z_{\dot\alpha})\,,
\end{equation}
where $\tilde\Psi( z_{\alpha},\bar z_{\dot\alpha})$ is the function on the two-dimensional complex affine plane parametrized by
two complex coordinates $ z_{\alpha} \in \ \stackrel{0}{\mathbb{C}}\!{}^2=\mathbb{C}^2\backslash(0,0)$.

Restoring the dependence of the wave function \eqref{wf-sol} on twistor variables using relations  \eqref{def-new-var-z} and \eqref{def-new-var-bz},
we obtain the twistor wave function in the form (we leave the notation for functions, as in \eqref{wf-sol})
\begin{equation}
\label{wf-tw}
\Psi( \pi_{\alpha},\bar \pi_{\dot\alpha};\rho_{\alpha},\bar\rho_{\dot\alpha}) \ = \
\delta\left(\pi^{\beta}\rho_{\beta}-M\right)\,\delta\left(\bar\rho_{\dot\beta}\bar\pi^{\dot\beta}-M\right)\,
e^{\displaystyle -i\left(\frac{\rho_{1}}{\pi_{1}}+\frac{\bar\rho_{1}}{\bar\pi_{1}}\right)}\,
\tilde\Psi( \pi_{\alpha},\bar \pi_{\dot\alpha})\,,
\end{equation}
where $\tilde\Psi( \pi_{\alpha},\bar \pi_{\dot\alpha})$ is an arbitrary function on $\pi_{\alpha}$, $\bar\pi_{\dot\alpha}$.
The twistor wave function, obtained in \eqref{wf-tw}, satisfies the following symmetry condition:
\begin{equation}
\label{sym-wf-tw}
\Psi ( \pi_{\alpha},\bar \pi_{\dot\alpha};\rho_{\alpha}+\kappa\,\pi_{\alpha},\bar\rho_{\dot\alpha}+\bar\kappa\,\bar\pi_{\dot\alpha}) \ = \
e^{\displaystyle -i\left(\kappa+\bar\kappa\right)}\,
\Psi ( \pi_{\alpha},\bar \pi_{\dot\alpha};\rho_{\alpha},\bar\rho_{\dot\alpha})\,,
\end{equation}
and, therefore, is the solution of  the equations
\footnote{Equations \eqref{equ-wf-tw} arise by differentiating expression \eqref{sym-wf-tw} with respect to $\kappa$, $\bar\kappa$
and the subsequent limits $\kappa=\bar\kappa=0$. }
\begin{equation}
\label{equ-wf-tw}
i\,\pi_{\alpha}\,\frac{\partial}{\partial\rho_{\alpha}}\,\Psi =  \Psi \,,\qquad
i\,\bar\pi_{\dot\alpha}\,\frac{\partial}{\partial\bar\rho_{\dot\alpha}}\,\Psi =  \Psi \,,
\end{equation}
which are quantum counterparts of the constraints  \eqref{constr-tw}.

Let us analyze spin (helicity) content of the twistor wave function \eqref{wf-tw}.
For it we find the value of
the Pauli-Lubanski operator  (see \eqref{PL-sp})
\begin{equation}
\label{PL-sp-op}
\mathbb{W}_{\alpha\dot\alpha}=
\bar {\mathbb{M}}_{\dot\alpha\dot\beta}\mathbb{P}^{\dot\beta}_{\alpha} - \mathbb{M}_{\alpha\beta}\mathbb{P}^{\beta}_{\dot\alpha}\,,
\end{equation}
where the Poincar\'{e} algebra operators in the twistor formulation have the form (see  \eqref{Lor-asamdual}, \eqref{M-tw})
\begin{eqnarray}
\mathbb{M}_{\alpha\beta} &=& \pi_{(\alpha}\,\frac{\partial}{\partial\pi^{\beta)}}+
\rho_{(\alpha}\,\frac{\partial}{\partial\rho^{\beta)}}\,,\nonumber\\
\bar{\mathbb{M}}_{\dot\alpha\dot\beta} &=& \bar\pi_{(\dot\alpha}\,\frac{\partial}{\partial\bar\pi^{\dot\beta)}}+
\bar\rho_{(\dot\alpha}\,\frac{\partial}{\partial\bar\rho^{\dot\beta)}}\,,\nonumber\\
\mathbb{P}_{\alpha\dot\alpha}&=& \pi_{\alpha} \bar\pi_{\dot\beta}  \,. \label{P-op}
\end{eqnarray}
Inserting  \eqref{P-op} in \eqref{PL-sp-op} we obtain
\begin{equation}
\label{PL-tw-op}
\mathbb{W}_{\alpha\dot\alpha}=
\pi_{\alpha}\bar\pi_{\dot\alpha}\,\mathbf{\Lambda}
+\frac{1}{2}\left[\pi_{\alpha}\bar\rho_{\dot\alpha} \Big(\bar\pi_{\dot\beta}\,\frac{\partial}{\partial\bar\rho_{\dot\beta}}\Big)
-\rho_{\alpha}\bar\pi_{\dot\alpha} \Big(\pi_{\beta}\,\frac{\partial}{\partial\rho_{\beta}}\Big)\right]
+\frac{1}{2}\left[(\bar\pi^{\dot\beta}\bar\rho_{\dot\beta})\pi_{\alpha}\,\frac{\partial}{\partial\bar\rho^{\dot\alpha}}- (\pi^{\beta}\rho_{\beta})\bar\pi_{\dot\alpha}\,\frac{\partial}{\partial\rho^{\alpha}}\right]\,,
\end{equation}
where
\begin{equation}
\label{hel-tw-op}
\mathbf{\Lambda}=-\frac{1}{2}\left(\pi_{\beta}\,\frac{\partial}{\partial\pi_{\beta}}-
\bar\pi_{\dot\beta}\,\frac{\partial}{\partial\bar\pi_{\dot\beta}}\right)\,.
\end{equation}
Direct calculations show that
\begin{equation}
\label{act-W-part}
\mathbb{W}_{\alpha\dot\alpha} \,\pi^{\beta}\rho_{\beta}=\mathbb{W}_{\alpha\dot\alpha} \,\bar\rho_{\dot\beta}\bar\pi^{\dot\beta}=0\,,\qquad
\mathbb{W}_{\alpha\dot\alpha} \left(\frac{\rho_{1}}{\pi_{1}}+\frac{\bar\rho_{1}}{\bar\pi_{1}}\right)=
\pi^{\beta}\rho_{\beta}\,\frac{\epsilon_{\alpha 1}\bar\pi_{\dot\alpha}}{\pi_1} \ - \
\bar\pi^{\dot\beta}\bar\rho_{\dot\beta}\,\frac{\epsilon_{\dot\alpha 1} \pi_{\alpha}}{\bar\pi_1}\,.
\end{equation}
By using \eqref{act-W-part} the action of the Pauli-Lubanski operator  \eqref{PL-tw-op} on the twistorial wave function  \eqref{wf-tw}
is the following:
\begin{equation}
\label{ct-W-wf-tw}
\mathbb{W}_{\alpha\dot\alpha} \Psi   =
\delta\left(\pi^{\beta}\rho_{\beta}-M\right)\delta\left(\bar\rho_{\dot\beta}\bar\pi^{\dot\beta}-M\right)
e^{\displaystyle -i\left(\frac{\rho_{1}}{\pi_{1}}+\frac{\bar\rho_{1}}{\bar\pi_{1}}\right)}
\,D_{\alpha\dot\alpha}\,\tilde\Psi \,.
\end{equation}
where the operator $D_{\alpha\dot\alpha}$, acting on the reduced twistor field $\tilde\Psi$, takes the form
\begin{equation}
\label{ct-D-wf-tw}
D_{\alpha\dot\alpha} \ :=
\ \pi_{\alpha}\bar\pi_{\dot\alpha}\,\mathbf{\Lambda}
+iM\left(\frac{\epsilon_{\dot\alpha 1} \pi_{\alpha}}{\bar\pi_1}
-\frac{\epsilon_{\alpha 1}\bar\pi_{\dot\alpha}}{\pi_1}\right)
\,.
\end{equation}
Acting on \eqref{ct-W-wf-tw} by the operator $\mathbb{W}^{\alpha\dot\alpha}$ and using
$D^{\alpha\dot\alpha}D_{\alpha\dot\alpha}\,\tilde\Psi=-2M^2\,\tilde\Psi$ we obtain that
(cf. \eqref{W2-sp})
\begin{equation}
\label{ct-W-wf-tw1}
\mathbb{W}^{\alpha\dot\alpha}\mathbb{W}_{\alpha\dot\alpha} \Psi  =
-2M^2\Psi_{tw}=-\mu^2\Psi\,.
\end{equation}
So the twistor field  \eqref{wf-tw} describes a massless particle of continuous spin.

The states with fixed helicities are the eigenvectors of the helicity operator $\mathbb{\Lambda}$ which is defined
as a projection of the total angular momentum $\vec{\mathbb{J}\,\,}$ in the direction of motion with momentum operator $\mathbb{P}_m=(\mathbb{P}_0,\vec{\mathbb{P}})$: $\mathbb{\Lambda}=\vec{\mathbb{J}\,\,}\vec{\mathbb{P}}/\mathbb{P}_0$.
In terms of the Pauli-Lubanski vector \eqref{PL-tw-op} this operator is defined by
\begin{equation}
\label{hel-exp}
\mathbb{\Lambda} \ = \ \frac{\mathbb{W}_{0}}{\mathbb{P}_{0}} \ = \
\frac{\mathbb{W}_{\alpha\dot\alpha}\tilde\sigma_0^{\dot\alpha\alpha}}{\pi_{\beta}\bar\pi_{\dot\beta}\tilde\sigma_0^{\dot\beta\beta}}\,.
\end{equation}
As we see from \eqref{ct-W-wf-tw} the representation \eqref{wf-tw} of the twistor field is unsuitable for helicity expansion
of the continuous spin wave function.

{}For finding helicity expansion of the twistor wave function  \eqref{wf-tw} we extract from $\tilde\Psi$ the exponential
multiplier:
\begin{equation}
\label{Psi-Psi}
\tilde\Psi( \pi_{\alpha},\bar \pi_{\dot\alpha}) \ = \
e^{\displaystyle \frac{-iM(\pi_{1}\pi_{2}+\bar\pi_{1}\bar\pi_{2})}{\pi_{1}\bar\pi_{1}(\pi_{\beta}\bar\pi_{\dot\beta}\tilde\sigma_0^{\dot\beta\beta})}}
\hat\Psi( \pi_{\alpha},\bar \pi_{\dot\alpha})\,.
\end{equation}
Inserting \eqref{Psi-Psi} in \eqref{wf-tw} we obtain the following representation of the twistorial field
\footnote{Here we take into account the fulfillment of equality
$$
\frac{\rho_{1}}{\pi_{1}}+\frac{\bar\rho_{1}}{\bar\pi_{1}}+\frac{\bar\pi^{\dot\alpha}\bar\rho_{\dot\alpha}\pi_{1}\pi_{2}+
\pi^{\alpha}\rho_{\alpha}\bar\pi_{1}\bar\pi_{2}}{\pi_{1}\bar\pi_{1}\sum\limits_{\beta=\dot\beta} \pi_{\beta}\bar\pi_{\dot\beta}} \ = \
\frac{\sum\limits_{\alpha=\dot\alpha} (\pi_{\alpha}\bar\rho_{\dot\alpha} + \rho_{\alpha}\bar\pi_{\dot\alpha})}
{\sum\limits_{\beta=\dot\beta} \pi_{\beta}\bar\pi_{\dot\beta}}\,.
$$
}
\begin{equation}
\label{wf-tw-hel}
\Psi ( \pi_{\alpha},\bar \pi_{\dot\alpha};\rho_{\alpha},\bar\rho_{\dot\alpha}) \ = \
\delta\left(\pi^{\beta}\rho_{\beta}-M\right)\,\delta\left(\bar\rho_{\dot\beta}\bar\pi^{\dot\beta}-M\right)\,
e^{\displaystyle \frac{-i(\pi_{\gamma}\bar\rho_{\dot\gamma}+\rho_{\gamma}\bar\pi_{\dot\gamma})\tilde\sigma_0^{\dot\gamma\gamma}}
{(\pi_{\delta}\bar\pi_{\dot\delta}\tilde\sigma_0^{\dot\delta\delta})}}\,
\hat\Psi ( \pi_{\alpha},\bar \pi_{\dot\alpha})\,.
\end{equation}
In the exponent of  \eqref{wf-tw-hel} there are in fact zero components of four-vectors \eqref{p-resol} and  \eqref{w-resol}.
So expression \eqref{wf-tw-hel} takes the form
\begin{equation}
\label{wf-tw-hel1}
\Psi ( \pi_{\alpha},\bar \pi_{\dot\alpha};\rho_{\alpha},\bar\rho_{\dot\alpha}) \ = \
\delta\left(\pi^{\beta}\rho_{\beta}-M\right)\,\delta\left(\bar\rho_{\dot\beta}\bar\pi^{\dot\beta}-M\right)\,
e^{\displaystyle -iw_0/p_0}\,
\hat\Psi ( \pi_{\alpha},\bar \pi_{\dot\alpha})\,,
\end{equation}
in which the quantities $p_0$ and $w_0$ have the resolved representations  \eqref{p-resol} and  \eqref{w-resol}.

The representation \eqref{wf-tw-hel1} for the twistorial field of the continuous spin particle is appropriate for
finding its helicity content. Taking into account the relation
\begin{eqnarray}
\label{act-W-part1}
\mathbb{W}_{\alpha\dot\alpha} \,\frac{w_{0}}{p_{0}}&=&
\frac{1}{2}\,\bigg[\pi_{\alpha}\bar\rho_{\dot\alpha} - \rho_{\alpha}\bar\pi_{\dot\alpha} -
\frac{\pi_{\alpha}\bar\pi_{\dot\alpha}}{\sum\limits_{\delta=\dot\delta}\pi_{\delta}\bar\pi_{\dot\delta}} \,
\sum\limits_{\beta=\dot\beta} (\pi_{\beta}\bar\rho_{\dot\beta} - \rho_{\beta}\bar\pi_{\dot\beta}) \bigg]\\
&&-\, \frac{1}{2\sum\limits_{\delta=\dot\delta}\pi_{\delta}\bar\pi_{\dot\delta}} \,
\bigg[(\bar\rho_{\dot\gamma}\bar\pi^{\dot\gamma})
\sum\limits_{\beta=\dot\beta} (\pi_{\alpha} \epsilon_{\dot\alpha\dot\beta}\pi_{\beta}) -
(\pi^{\gamma} \rho_{\gamma}) \sum\limits_{\beta=\dot\beta} (\bar\pi_{\dot\alpha}\epsilon_{\alpha\beta} \bar\pi_{\dot\beta})\bigg]\,, \nonumber
\end{eqnarray}
which holds for the operator \eqref{PL-tw-op} and resolved representations  \eqref{p-resol}, \eqref{w-resol} for $p_0$, $w_0$,
we obtain that the action of the Pauli-Lubanski vector on the field \eqref{wf-tw-hel1} takes the form
\begin{eqnarray}
\label{ct-W-wf-tw-h}
\mathbb{W}_{\alpha\dot\alpha} \Psi   &=&
\delta\left(\pi^{\beta}\rho_{\beta}-M\right)\delta\left(\bar\rho_{\dot\beta}\bar\pi^{\dot\beta}-M\right)
e^{\displaystyle -iw_0/p_0}
\, \hat D_{\alpha\dot\alpha}\,\hat\Psi   \,,
\end{eqnarray}
where
\begin{equation}
\label{D-new}
\begin{array}{rcl}
\hat D_{\alpha\dot\alpha}&=&{ \displaystyle \pi_{\alpha}\bar\pi_{\dot\alpha}\,\mathbf{\Lambda}
- \frac{i}{2}\,\big(\pi_{\alpha}\bar\rho_{\dot\alpha}  -  \rho_{\alpha}\bar\pi_{\dot\alpha}\big)}
\\ [7pt]
&& { \displaystyle  + \
\frac{i\pi_{\alpha}\bar\pi_{\dot\alpha}}{2\sum\limits_{\delta=\dot\delta}\pi_{\delta}\bar\pi_{\dot\delta}} \,
\sum\limits_{\beta=\dot\beta} (\pi_{\beta}\bar\rho_{\dot\beta} - \rho_{\beta}\bar\pi_{\dot\beta})   +
\frac{iM}{2\sum\limits_{\delta=\dot\delta}\pi_{\delta}\bar\pi_{\dot\delta}} \,
\sum\limits_{\beta=\dot\beta} (\pi_{\alpha} \epsilon_{\dot\alpha\dot\beta}\pi_{\beta}-
\bar\pi_{\dot\alpha}\epsilon_{\alpha\beta} \bar\pi_{\dot\beta})
\,.}
\end{array}
\end{equation}
In contrast to the quantity \eqref{ct-D-wf-tw}, the operator  \eqref{D-new} satisfies the property
$$
\sum\limits_{\alpha=\dot\alpha}\hat D_{\alpha\dot\alpha}=\sum\limits_{\alpha=\dot\alpha}\pi_{\alpha}\bar\pi_{\dot\alpha}\,\mathbf{\Lambda}=
\mathbb{P}_{0}\mathbf{\Lambda}\,.
$$
As a result, the helicity operator \eqref{hel-exp} acts on the twistorial field in the following way:
\begin{equation}
\label{hel-act-wf}
\mathbb{\Lambda} \, \Psi \ = \ \frac{\mathbb{W}_{0}}{\mathbb{P}_{0}}\, \Psi \ = \
\frac{\sum\limits_{\alpha=\dot\alpha}\mathbb{W}_{\alpha\dot\alpha}}{\sum\limits_{\beta=\dot\beta}\pi_{\beta}\bar\pi_{\dot\beta}} \, \Psi
\ = \ \delta\left(\pi^{\beta}\rho_{\beta}-M\right)\delta\left(\bar\rho_{\dot\beta}\bar\pi^{\dot\beta}-M\right)
e^{\displaystyle -iw_0/p_0}
\,\mathbf{\Lambda} \hat\Psi \,.
\end{equation}
As we see from this expression,
the eigenvalues of the helicity operator $\mathbb{\Lambda}$ is defined by the action of the operator $\mathbf{\Lambda}$ on the reduced twistorial function
$\hat\Psi ( \pi,\bar \pi)$ living
on the two-dimensional complex affine plane parametrized by $\pi_{\alpha}$.

The eigenvectors of the operator $\mathbf{\Lambda}$ are homogeneous functions of two complex variables.
For this reason we express the function $\hat\Psi ( \pi,\bar \pi)$
by its homogeneous components.

The homogeneous components of the function $\hat\Psi ( \pi,\bar \pi)$
are determined by using the Mellin transform \cite{GelGV}
\begin{equation}
\label{Mel-trans}
F^{(n_1,n_2)}( \pi_{\alpha},\bar \pi_{\dot\alpha}) \ = \ \frac{i}{2}\int d\lambda\, d\bar\lambda\,
\,\lambda^{-n_1\,}\bar\lambda^{-n_2}\,\hat\Psi ( \lambda\,\pi_{\alpha},\bar\lambda\,\bar \pi_{\dot\alpha})\,,
\end{equation}
where $\chi=(n_1,n_2)$ are the pair of complex numbers  whose difference is equal to an integer number.
The functions \eqref{Mel-trans} are homogeneous of bi-degree $(n_1-1,n_2-1)$:
\begin{equation}
\label{Mel-trans-hom}
F^{(n_1,n_2)}( a\pi_{\alpha},\bar a\bar \pi_{\dot\alpha}) \ = \
a^{n_1-1}\bar a^{n_2-1}F^{(n_1,n_2)}( \pi_{\alpha},\bar \pi_{\dot\alpha}) \,.
\end{equation}
In the expansion of quadratically integrable functions on the two-dimensional complex affine plane,
in whose space the unitary representation of the Lorentz group is realized,
there are only such homogeneous functions $F^{(\chi)}=F^{(n_1,n_2)}$ in which
${\displaystyle \chi=\left(\frac{n+i\nu}{2},\frac{-n+i\nu}{2}\right)}$ where $n$ is integer and $\nu$ is an arbitrary real number \cite{GelGV}.
On this space
of the homogeneous functions of bi-degree ${\displaystyle \chi=\left(\frac{n+i\nu}{2},\frac{-n+i\nu}{2}\right)}$
the principal series of unitary irreducible representations of Lorentz group $\mathrm{SL}(2,\mathbb{C})$ is realized.
Inverse transformation for \eqref{Mel-trans} has the form \cite{GelGV}
\begin{equation}
\label{invers-Mel-trans}
\hat\Psi ( \pi_{\alpha},\bar \pi_{\dot\alpha}) \ = \
\frac{1}{4\pi^2}\sum\limits_{n=-\infty}^{\infty} \int\limits_{-\infty}^{\infty} d\nu\,
F^{\textstyle \left(\frac{n+i\nu}{2},\frac{-n+i\nu}{2}\right)}
( \pi_{\alpha},\bar \pi_{\dot\alpha})\,.
\end{equation}

The condition \eqref{Mel-trans-hom} is equivalent to the fulfillment of the equations
\begin{equation}
\label{Mel-trans-hom-eqs}
\pi_{\alpha}\frac{\partial}{\partial\pi_{\alpha}}\,F^{(n_1,n_2)} =
(n_1-1)F^{(n_1,n_2)}\,,\qquad
\bar\pi_{\dot\alpha}\frac{\partial}{\partial\bar\pi_{\dot\alpha}}\,F^{(n_1,n_2)} =
(n_2-1)F^{(n_1,n_2)} \,.
\end{equation}
On such components, the helicity operator \eqref{hel-tw-op} takes the values $s=-n/2$:
\begin{equation}
\label{Mel-trans-hom-hel}
\mathbf{\Lambda}\,F^{(n_1,n_2)} =
-\frac{n}{2}\,F^{(n_1,n_2)} \,.
\end{equation}
As a result this and \eqref{hel-act-wf}, the twistorial wave function \eqref{wf-tw-hel1}
of the continuous spin particle describes an infinite number of massless states
whose helicities are equal to integer or half-integer values and run from $-\infty$ to $+\infty$.
Expression \eqref{invers-Mel-trans} can be interpreted as expansion on the helicity states
\begin{equation}
\label{exp-hel-Mel-trans}
\hat\Psi ( \pi_{\alpha},\bar \pi_{\dot\alpha}) \ = \
\sum\limits_{n=-\infty}^{\infty} \mathcal{F}^{(n)}( \pi_{\alpha},\bar \pi_{\dot\alpha})
\end{equation}
after introducing the fields
\begin{equation}
\label{hel-comp}
\mathcal{F}^{(n)}( \pi_{\alpha},\bar \pi_{\dot\alpha}) \ = \
\frac{1}{4\pi^2} \int\limits_{-\infty}^{\infty} d\nu\,
F^{\textstyle \left(\frac{n+i\nu}{2},\frac{-n+i\nu}{2}\right)}
( \pi_{\alpha},\bar \pi_{\dot\alpha})\,.
\end{equation}
Precisely these twistorial fields $\mathcal{F}^{(n)}$ describe massless states with helicities $s=-n/2$.

Imposing the parity conditions to the twistor field
\eqref{wf-tw-hel1}
\begin{equation}
\label{wf-tw-par}
\Psi ( \pi_{\alpha},\bar \pi_{\dot\alpha};\rho_{\alpha},\bar\rho_{\dot\alpha}) \ = \
(-1)^{\nu}\Psi ( -\pi_{\alpha},-\bar \pi_{\dot\alpha};-\rho_{\alpha},-\bar\rho_{\dot\alpha})
\end{equation}
and also
\begin{equation}
\label{wf-tw-red-par}
\hat\Psi ( \pi_{\alpha},\bar \pi_{\dot\alpha}) \ = \
(-1)^{\nu}\hat\Psi ( -\pi_{\alpha},-\bar \pi_{\dot\alpha})\,,
\end{equation}
where $\nu=0$ or $\nu=1$, we then find that in the expansion \eqref{exp-hel-Mel-trans} there are only terms with
even values of $n$ at $\nu=0$ or only terms with
odd values of $n$ at $\nu=1$.
Thus, even twistor fields \eqref{wf-tw-hel1} describe an infinite number of massless states with integer helicities,
whereas odd twistor fields are used to determine massless states with half-integer helicities.

\setcounter{equation}{0}
\section{Twistor transform for fields of continuous spin particles}

\quad\, In the previous Section, we obtained the twistor fields describing massless particles of continuous spin.
Here we will establish the connection of these twistor fields with the corresponding fields of the space-time formulation.
These space-time fields will depend on the position four-vector and obey the Wigner-Bargmann equations \cite{Wigner47,BargWigner}
which are quantum counterparts of the constraints \eqref{const-sp}-\eqref{const-sp-3}.

The twistorial variables $\pi_{\alpha}$, $\rho_{\alpha}$, on which the wave function \eqref{wf-tw} depends, play a role
of momentum variables.
Therefore, we can consider the twistor wave function \eqref{wf-tw} as a wave function in the momentum space.
For this reason, up to the normalization multiplier the space-time wave function can be determined by means of the integral transformation
\begin{equation}
\label{wf-st-tw}
\Phi(x,y) \ = \ \int d^4 \pi \, d^4 \rho \, e^{\displaystyle i p_{\alpha\dot\alpha} x^{\dot\alpha\alpha}}\,
e^{\displaystyle i w_{\alpha\dot\alpha} y^{\dot\alpha\alpha}}\, \Psi (\pi,\bar\pi;\rho,\bar\rho)\,,
\end{equation}
where the momenta $p_{\alpha\dot\alpha}$ and $w_{\alpha\dot\alpha}$ are composite and
defined by expressions \eqref{p-resol}, \eqref{w-resol}:
$p_{\alpha\dot\alpha}=\pi_{\alpha}\bar\pi_{\dot\alpha}$,
$w_{\alpha\dot\alpha}=\pi_{\alpha}\bar\rho_{\dot\alpha}+\rho_{\alpha}\bar\pi_{\dot\alpha}$.
In the integral \eqref{wf-st-tw} we perform integration over the four-dimensional complex space with the integration measure
$d^4 \pi \, d^4 \rho := \frac14\, d\pi^{\alpha}\wedge d\pi_{\alpha}\wedge d\bar\pi_{\dot\alpha}\wedge d\bar\pi^{\dot\alpha}
\wedge \frac14\, d\rho^{\beta}\wedge d\rho_{\beta}\wedge d\bar\rho_{\dot\beta}\wedge d\bar\rho^{\dot\beta}$.

Inserting \eqref{wf-tw} in   \eqref{wf-st-tw} we obtain the space-time fields of continuous spin particles
\begin{equation}
\label{wf-st-det}
\!\!\! \Phi(x,y) =\! \int \! d^4 \pi \, d^4 \rho \, e^{\displaystyle i p_{\alpha\dot\alpha} x^{\dot\alpha\alpha}}
e^{\displaystyle i w_{\alpha\dot\alpha} y^{\dot\alpha\alpha}}
\delta\left(\pi^{\beta}\rho_{\beta}-M\right)\delta\left(\bar\rho_{\dot\beta}\bar\pi^{\dot\beta}-M\right)
e^{\displaystyle -i\left(\frac{\rho_{1}}{\pi_{1}}+\frac{\bar\rho_{1}}{\bar\pi_{1}}\right)}
\tilde\Psi ( \pi,\bar \pi)\,.
\end{equation}
Integral representations \eqref{wf-st-det} for the space-time fields are the solution of the Wigner-Bargmann equations \cite{Wigner47,BargWigner},
which are quantum counterparts of the constraints \eqref{const-sp}-\eqref{const-sp-3}.
The presence of equations \eqref{eq-sp}-\eqref{eq-sp-2} for the fields \eqref{wf-st-det} is the consequence of the identities
$\pi^\alpha\pi_\alpha\equiv 0$, $\rho^\alpha\rho_\alpha\equiv 0$ for even Weyl spinors and
the conditions \eqref{M-constr-def} encoded by means of the $\delta$-functions in the integrand \eqref{wf-st-det}.
The proof of the fulfillment of equation \eqref{eq-sp-3} is the following:
\begin{eqnarray}\label{eq-wf-ful}
&&-\,i\,y^{\dot\alpha\alpha}\,\frac{\partial}{\partial x^{\dot\alpha\alpha}} \ \Phi = \\
&&\! \int \! d^4 \pi  d^4 \rho \, e^{\displaystyle i p_{\alpha\dot\alpha} x^{\dot\alpha\alpha}}
\pi_{\alpha}\bar\pi_{\dot\alpha} y^{\dot\alpha\alpha}
e^{\displaystyle i w_{\alpha\dot\alpha} y^{\dot\alpha\alpha}} \delta\left(\pi^{\beta}\rho_{\beta}-M\right)\delta\left(\bar\rho_{\dot\beta}\bar\pi^{\dot\beta}-M\right)
e^{\displaystyle -i\left(\frac{\rho_{1}}{\pi_{1}}+\frac{\bar\rho_{1}}{\bar\pi_{1}}\right)}
\tilde\Psi   = \nonumber\\
&&\! -i\int \! d^4 \pi  d^4 \rho \, e^{\displaystyle i p_{\alpha\dot\alpha} x^{\dot\alpha\alpha}}
\left[\pi_\alpha\frac{\partial}{\partial \rho_\alpha}\,e^{\displaystyle i w_{\alpha\dot\alpha} y^{\dot\alpha\alpha}} \right] \delta\left(\pi^{\beta}\rho_{\beta}-M\right)\delta\left(\bar\rho_{\dot\beta}\bar\pi^{\dot\beta}-M\right)
e^{\displaystyle -i\left(\frac{\rho_{1}}{\pi_{1}}+\frac{\bar\rho_{1}}{\bar\pi_{1}}\right)}
\tilde\Psi   = \nonumber\\
&&\! i\!\int \! d^4 \pi  d^4 \rho \, e^{\displaystyle i p_{\alpha\dot\alpha} x^{\dot\alpha\alpha}}
e^{\displaystyle i w_{\alpha\dot\alpha} y^{\dot\alpha\alpha}}
\delta\left(\pi^{\beta}\rho_{\beta}-M\right)\delta\left(\bar\rho_{\dot\beta}\bar\pi^{\dot\beta}-M\right)
\bigg[\pi_\alpha\frac{\partial}{\partial \rho_\alpha}\,e^{\displaystyle -i\left(\frac{\rho_{1}}{\pi_{1}}+\frac{\bar\rho_{1}}{\bar\pi_{1}}\right)}\bigg]\,
\tilde\Psi   =  \Phi\,. \nonumber
\end{eqnarray}

Thus, we have constructed the integral relationship \eqref{wf-st-det}
between the space-time fields and twistor ones,
which is a generalization of the Penrose field twistor transform.
The twistor function $\tilde\Psi ( \pi,\bar\pi)$ plays the role of the prepotential
for the space-time field $\Phi$.

In expression  \eqref{wf-st-det} we can perform the integration over $\rho$, after which only the integrals over the $\pi$-components remain.
It should be noted that relations \eqref{def-new-var-z} and \eqref{def-new-var-bz} imply
\begin{equation}
\label{pi-stlambda}
\rho_{\alpha}=t\pi_{\alpha} +\frac{Ms}{\pi_1}\,\delta_\alpha^2\,,\qquad
\bar\rho_{\dot\alpha}=\bar t\bar\pi_{\dot\alpha} +\frac{M\bar s}{\bar\pi_1}\,\delta_{\dot\alpha}^2
\end{equation}
and, therefore,
$d^4 \pi \, d^4 \rho = \frac14\, M^2 d\pi^{\alpha}\wedge d\pi_{\alpha}\wedge d\bar\pi_{\dot\alpha}\wedge d\bar\pi^{\dot\alpha}
\wedge ds\wedge d\bar s\wedge dt\wedge d\bar t$.
Integration with respect to $s$, $\bar s$ is performed by means of the $\delta$-functions.
Integration with respect to $t+\bar t$ produces $\delta(\pi_{\alpha}\bar\pi_{\dot\alpha}y^{\dot\alpha\alpha}-1)$
and integration with respect to $i(t-\bar t)$  is factorized. So expression  \eqref{wf-st-det}
for the field of the continuous spin particle takes, up to the multiplier constant, the form
\begin{equation}
\label{wf-st-det-1}
\Phi(x,y) = \int \! d^4 \pi \, e^{\displaystyle i \pi_{\alpha}\bar\pi_{\dot\alpha} x^{\dot\alpha\alpha}}\
e^{\displaystyle iM\left(\frac{\pi_{\alpha}\delta_{\dot\alpha}^2}{\pi_{1}}+
\frac{\bar\pi_{\dot\alpha}\delta_{\alpha}^2}{\bar\pi_{1}}\right)y^{\dot\alpha\alpha}}
\delta(\pi_{\alpha}\bar\pi_{\dot\alpha}y^{\dot\alpha\alpha}-1)\
\tilde\Psi ( \pi,\bar\pi)\,.
\end{equation}
We remind that $2M^2=\mu^2$. It is easy to check that the  field
\eqref{wf-st-det-1} satisfies equations
\eqref{eq-sp}-\eqref{eq-sp-3}. Therefore, it is the exact solution of
these equations in the twistorial representation. It is evident that
at $\mu=0$ we obtain the exact solution to equations
\eqref{eq-sp}-\eqref{eq-sp-3} at $\mu=0$.

\setcounter{equation}{0}
\section{Summary and outlook}

\quad\,
We constructed the new Lagrangian model which describes a relativistic massless particle of the continuous spin.
This model is characterized by the following:
\begin{itemize}
\item The classical Lagrangian \eqref{L-sp} describes the relativistic particle corresponding to the irreducible
massless representation of the Poincar\'{e} group with continuous spin.
\item The classical twistor Lagrangian \eqref{L-tw} with twistor constraints \eqref{M-constr-def}, \eqref{constr-tw}, \eqref{const-tw-4} and
coordinate twistor transform \eqref{p-resol}, \eqref{w-resol}, \eqref{y-tw}, \eqref{z-tw} which give the links
between phase variables of space-time formulation and twistors.
\item Twistor fields of the continuous spin massless particles \eqref{wf-tw-hel} (or \eqref{wf-tw})
which depend only on twistor variables and have the helicities expansion \eqref{exp-hel-Mel-trans}.
\item Field twistor transform \eqref{wf-st-det} which expresses space-time fields in terms of twistor fields by the integral transformation.
\item Space-time field \eqref{wf-st-det-1} is the exact solution of the constraints \eqref{eq-sp}-\eqref{eq-sp-3}.
\end{itemize}

Let us note some comments on the constructed model.
\begin{description}
\item[i)]
Mass parameter $\mu$ determining the irreducible Poincar\'{e} representations has a role similar
to the mass in the massive Poincar\'{e} representations: spin (helicity) contents of
the continuous spin representations are the same for different values of $\mu$.
The dependence on $\mu$ is coded in the explicit dependence of the twistor fields on the twistor variables
and, consequently, in the equations for the space-time fields.

\item[ii)]
In the limit $\mu\to 0$ our twistorial model produces the massless higher spin particle.
So in the mixed space-time--twistorial formulation a higher spin particle is described by the even coordinates,
four-vector $x^m$ and the Weyl spinor $\pi_\alpha$, and its canonically-conjugated momenta, which are subjected
to the constraint (see \eqref{p-resol})
\begin{equation}
\label{hsp-constr}
p_{\alpha\dot\alpha}-\pi_{\alpha}\bar\pi_{\dot\alpha}\approx0\,.
\end{equation}
This constraint yields after quantization the unfolded equation for a free higher spin field \cite{V+} (see also \cite{FI})
in the representation with momentum realization for $\pi$-variables.
In the coordinate representation  for the Weyl spinor $\pi$ the solution of the constraint \eqref{hsp-constr} is the following
field:
\begin{equation}
\label{hsp-field}
\Psi_{hsp}(x,\pi,\bar\pi)=e^{\displaystyle i \pi_{\alpha}\bar\pi_{\dot\alpha}x^{\dot\alpha\alpha}}
\tilde\Psi_{hsp}(\pi,\bar\pi)\,,
\end{equation}
where the function $\tilde\Psi_{hsp}(\pi,\bar\pi)$ of two complex variables $\pi_\alpha$ defines
an infinite tower of the massless states with arbitrary helicities.
However, the twistor wave function \eqref{wf-tw} of the continuous spin particle is defined also by a similar
function $\tilde\Psi ( \pi,\bar \pi)$. Moreover, in the limit $\mu\to 0$ ($M\to 0$)
the constraints $\pi^{\alpha}\rho_{\alpha}=M=0$ and c.c. imply $\pi_{\alpha}\sim\rho_{\alpha}$.
Therefore, in this case we have $w_{\alpha\dot\alpha}\sim p_{\alpha\dot\alpha}$ due to \eqref{w-resol} and
the integrand in \eqref{wf-st-det} coincides with \eqref{hsp-field} after considering a linear combination of
two 4-vectors $x^m$ and $y^m$ as some space-time position four-vector.
This gives us an additional relation between the continuous spin representations and other studied theories,
in addition to their known link with the contraction of the finite spin representations in higher space-time dimensions
(see, for example, \cite{BekMou,BekSk}).

\item[iii)]
There is a mixed Shirafuji formulation \cite{S83} of massless particles which uses both space-time vector variables
and twistorial ones, with a specific term $\pi_{\alpha}\bar\pi_{\dot\alpha}\dot x^{\dot\alpha\alpha}$ in the Lagrangian.
This formulation of the continuous spin particle will make it possible to find additional relations of this system
with a higher spin particle in the unfolded formulation.
In particular, path-integral quantization of such models will yield an expression for the
transition amplitudes of the continuous spin particles.
In our future research, we will study this issue.

\item[iv)]
Another important task in this subject is the construction of the Lagrangian field
(second-quantized in the context of the current work) theory of continuous spin.
An effective way of realizing this task is to use BRST quantization methods.
Some aspects of BRST formulations of the field theories in the case of continuous spin particles were studied in
\cite{Bengtsson13,Metsaev18,AlkGr}.
In our forthcoming work, we will study the BRST-BFV formulation for the twistorial continuous spin particle constructed here.
The first natural step in this direction will be the study of the mixed Shirafuji-like model
in which twistor variables will play the role of some creation and annihilation operators in the the continuous spin particle system,
similar to the studies carried out in~\cite{BuchKrP,BuchKr}  (see also the recent preprint
\cite{BuchKrTak}  and references therein).

\item[v)]
On the (A)dS spaces, one can also consider continuous-spin fields
describing infinite-dimensional irreducible representations of the (A)dS groups.
Such representations
were studied in the papers~\cite{Mets16,Mets17}  (see also the review \cite{BekSk}).
The (A)dS spaces play a special role in constructing the interaction of high-spin fields with a gravitational field.
Therefore, it is interesting to generalize the twistor formalism proposed here to the theory of continuous spin particles
in the (A)dS spaces with a gravitational background.
\end{description}

\section*{Acknowledgments}
The research of I.L.B. \& S.F. was supported by
the Russian Ministry of Education and Science, project No.\,3.1386.2017 and
the Russian Science Foundation, grant No.\,16-12-10306.
I.L.B. acknowledges also the support of the  Russian Foundation for Basic Research, project No.\,18-02-00153.
A.P.I. acknowledges the support of the Russian Science Foundation, grant No.\,14-11-00598.


\begin{thebibliography}{96}

\bibitem{Wigner39}
E.P.\,Wigner,
{\it On unitary representations of the inhomogeneous Lorentz group},
Annals Math.  {\bf 40} (1939) 149.

\bibitem{Wigner47}
E.P.\,Wigner,
{\it Relativistische Wellengleichungen},
Z. Physik  {\bf 124} (1947) 665.

\bibitem{BargWigner}
V.\,Bargmann, E.P.\,Wigner,
{\it Group theoretical discussion of relativistic wave equations},
Proc. Nat. Acad. Sci. US  {\bf 34} (1948) 211.


\bibitem{Iv-Mack}
G.J.\,Iverson, G.\,Mack,
{\it Quantum fields and interactions of massless particles - the continuous spin case},
Annals Phys.  {\bf 64} (1971) 253.


\bibitem{Vas1989}
M.A.\,Vasiliev,
{\it Consistent equations for interacting massless fields of all spins in the first order in curvatures},
Annals Phys. {\bf 190} (1972) (1989) 59.

\bibitem{Vas1991}
M.A.\,Vasiliev,
{\it Algebraic  aspects  of  the  higher  spin  problem},
Phys. Lett. {\bf B257} (1991) 111.


\bibitem{Vas1992}
M.A.\,Vasiliev,
{\it More on equations of motion for interacting massless fields of all spins in (3+1)-dimensions},
Phys. Lett. {\bf B285} (1992) 225.


\bibitem{Vas2001}
M.A.\,Vasiliev,
{\it Progress in higher spin gauge theories},
{Proceedings of the International Conference on Quantization, Gauge Theory, and Strings:
Conference Dedicated to the Memory of Prof. E.\,Fradkin,}
Eds. A.\,Semikhatov, M.\,Vasiliev, V.\,Zaikin, Scientific World, Moscow, 2001, 452-472, {\tt arXiv: hep-th/0104246}.


\bibitem{Vas2002}
M.A.\,Vasiliev,
{\it Relativity, causality, locality, quantization and duality in the $Sp(2M)$ invariant generalized spacetime},
in {Multiple Facets of Quantization and Supersymmetry}, Michael Marinov Memorial Volume,
Eds. M.\,Olshanetsky and A.\,Vainshtein, World Scientific, 2002,
826-872,
{\tt arXiv:hep-th/0111119}.

\bibitem{Vas2005}
X.\,Bekaert, S.\,Cnockaert, C.\,Iazeolla, M.A.\,Vasiliev,
{\it Nonlinear higher spin theories in various dimensions},
{Proceedings of the 1st Solvay Workshop on Higher Spin Gauge Theories, 12-14 May 2004. Brussels, Belgium},
Eds. R.\,Argurio, G.\,Barnich, G.\,Bonelli, M.\,Grigoriev, Int. Solvay Institutes, 2006,
132-197,
{\tt arXiv:hep-th/0503128}.


\bibitem{Vas2018}
M.A.\,Vasiliev,
{\it From Coxeter Higher-Spin Theories to Strings and Tensor Models},
{\tt arXiv:1804.06520\,[hep-th]}.

\bibitem{MundSY}
J.\,Mund, B.\,Schroer, J.\,Yngvason,
{\it String localized quantum fields from Wigner representations},
Phys. Lett.  {\bf B596} (2004) 156, {\tt arXiv:math-ph/0402043}.


\bibitem{BKRX}
L.\,Brink, A.M.\,Khan, P.\,Ramond, X.-Z. Xiong,
{\it Continuous spin representations of the Poincar\'{e} and superPoincar\'{e} groups},
J. Math. Phys.  {\bf 43} (2002) 6279, {\tt arXiv:hep-th/0205145}.

\bibitem{BekBoul}
X.\,Bekaert, N.\,Boulanger,
{\it The unitary representations of the Poincar\'{e} group in any spacetime dimension},
Lectures presented at
2nd Modave Summer School in Theoretical Physics,
6-12 Aug 2006, Modave, Belgium, {\tt arXiv:hep-th/0611263}.



\bibitem{BekMou}
X.\,Bekaert, J.\,Mourad,
{\it The continuous spin limit of higher spin field equations},
JHEP  {\bf 0601} (2006) 115, {\tt arXiv:hep-th/0509092}.

\bibitem{SchToro13a}
P.\,Schuster, N.\,Toro,
{\it On the Theory of Continuous-Spin Particles: Wavefunctions and Soft-Factor Scattering Amplitudes},
JHEP  {\bf 1309} (2013) 104, {\tt arXiv:1302.1198\,[hep-th]}.

\bibitem{SchToro13b}
P.\,Schuster, N.\,Toro,
{\it On the Theory of Continuous-Spin Particles: Helicity Correspondence in Radiation and Forces},
JHEP  {\bf 1309} (2013) 105, {\tt arXiv:1302.1577\,[hep-th]}.

\bibitem{SchToro13c}
P.\,Schuster, N.\,Toro,
{\it A Gauge Field Theory of Continuous-Spin Particles},
JHEP  {\bf 1310} (2013) 061, {\tt arXiv:1302.3225\,[hep-th]}.


\bibitem{SchToro15}
P.\,Schuster, N.\,Toro,
{\it A CSP Field Theory with Helicity Correspondence},
Phys. Rev.  {\bf D91} (2015) 025023, {\tt arXiv:1404.0675\,[hep-th]}.


\bibitem{Riv}
V.O.\,Rivelles,
{\it Gauge theory formulations for continuous and higher spin fields},
Phys. Rev.  {\bf D91} (2015) 125035, {\tt arXiv:1408.3576\,[hep-th]}.

\bibitem{Mets16}
R.R.\,Metsaev,
{\it Continuous spin gauge field in (A)dS space},
Phys. Lett.    {\bf B767} (2017) 458, {\tt arXiv:1610.00657\,[hep-th]}.

\bibitem{Mets17}
R.R.\,Metsaev,
{\it Fermionic continuous spin gauge field in (A)dS space},
Phys. Lett.    {\bf B773} (2017) 135, {\tt arXiv:1703.05780\,[hep-th]}.


\bibitem{BekSk}
X.\,Bekaert, E.D.\,Skvortsov,
{\it Elementary particles with continuous spin},
Int. J. Mod. Phys.   {\bf A32} (2017) 1730019, {\tt arXiv:1708.01030\,[hep-th]}.

\bibitem{HabZin}
M.V.\,Khabarov, Yu.M.\,Zinoviev,
{\it Infinite (continuous) spin fields in the frame-like formalism},
Nucl. Phys. {\bf B928} (2018) 182, {\tt arXiv:1711.08223\,[hep-th]}.

\bibitem{AlkGr}
K.B.\,Alkalaev, M.A.\,Grigoriev,
{\it Continuous spin fields of mixed-symmetry type},
JHEP {\bf 1803} (2018) 030, {\tt arXiv:1712.02317\,[hep-th]}.

\bibitem{BK}
I.L.\,Buchbider, S.M.\,Kuzenko, {\it Ideas and Methods of Supersymmetry
and Supergravity}, IOP Publ., 1998.

\bibitem{Pen67}
R.\,Penrose,
{\it Twistor algebra},
J. Math. Phys. {\bf 8} (1967) 345.

\bibitem{PC72}
R.\,Penrose, M.A.H.\,MacCallum,
{\it Twistor theory: an approach to the quantization of fields and spacetime},
Phys. Rept. {\bf 6} (1972) 241.


\bibitem{PenRin}
R.\,Penrose, W.\,Rindler,
{\it Spinors And Space-time. Vol. 2: Spinor And Twistor Methods In Space-time Geometry},
Cambridge University Press, 1988, 512 pages.


\bibitem{BL99}
I.\,Bandos, J.\,Lukierski,
{\it Tensorial central charges and new superparticle models with fundamental spinor coordinates},
Mod. Phys. Lett. {\bf A14} (1999) 1257, {\tt arXiv:hep-th/9811022}.

\bibitem{BLS00}
I.\,Bandos, J.\,Lukierski, D.\,Sorokin,
{\it Superparticle models with tensorial central charges},
Phys. Rev. {\bf D61}, 045002 (2000), {\tt arXiv:hep-th/9904109}.

\bibitem{V+}
M.A.\,Vasiliev,
{\it Conformal higher spin symmetries of 4d massless supermultiplets and $osp(L,2M)$ invariant equations
in generalized (super)space},
Phys. Rev. {\bf D66} (2002) 066006, {\tt arXiv:hep-th/0106149}.

\bibitem{FI}
S.\,Fedoruk, E.\,Ivanov,
{\it Master higher-spin particle},
Class. Quant. Grav. {\bf 23} (2006) 5195, {\tt arXiv:hep-th/0604111}.


\bibitem{FZ03}
S.\,Fedoruk, V.G.\,Zima,
{\it Bitwistor formulation of massive spinning particle},
Journal of Kharkov University {\bf 585} (2003) 39, {\tt arXiv:hep-th/0308154}.

\bibitem{FFLM06}
S.\,Fedoruk, A.\,Frydryszak, J.\,Lukierski, C.\,Miquel-Espanya,
{\it Extension of the Shirafuji model for massive particles with spin},
Int. J. Mod. Phys. {\bf A21} (2006) 4137, {\tt arXiv:hep-th/0510266}.

\bibitem{FL14}
S.\,Fedoruk, J.\,Lukierski,
{\it Massive twistor particle with spin generated by Souriau-Wess-Zumino term and its quantization},
Phys. Lett. {\bf B733} (2014) 309,
{\tt arXiv:1403.4127\,[hep-th]}.

\bibitem{IsP}
A.P.\,Isaev, M.A.\,Podoinitsyn,
{\it Two-spinor description of massive particles and relativistic spin projection operators},
Nucl. Phys.  {\bf B929} (2018) 452,
{\tt arXiv:1712.00833\,[hep-th]}.

\bibitem{AFIL}
J.A.\,de Azcarraga, S.\,Fedoruk, J.M.\,Izquierdo, J.\,Lukierski,
{\it Two-twistor particle models and free massive higher spin fields},
JHEP  {\bf 1504} (2015) 010,
{\tt arXiv:1409.7169\,[hep-th]}.

\bibitem{Band}
I.A.\,Bandos, {\it Superparticle in Lorentz harmonic superspace} (in Russian),
Sov. J. Nucl. Phys. {\bf 51} (1990) 906.

\bibitem{FedZim}
S.\,Fedoruk, V.G.\,Zima,
{\it Covariant quantization of d=4 Brink-Schwarz superparticle with Lorentz harmonics},
Theor. Math. Phys. {\bf 102} (1995) 305, {\tt arXiv:hep-th/9409117}.


\bibitem{GelGV}
I.M.\,Gelfand, M.I.\,Graev, N.J.\,Vilenkin,
{\it Generalized Functions: Integral geometry and representation theory},
New York, London: Academic Press, 1966.



\bibitem{S83}
T.\,Shirafuji,
{\it Lagrangian mechanics of massless particles with spin},
Prog. Theor. Phys. {\bf 70} (1983) 18.


\bibitem{Bengtsson13}
A.K.H.\,Bengtsson,
{\it BRST Theory for Continuous Spin},
JHEP {\bf 1310} (2013) 108,
{\tt arXiv:1303.3799\,[hep-th]}.

\bibitem{Metsaev18}
R.R.\,Metsaev,
{\it BRST-BV approach to continuous-spin field},
Phys. Lett. {\bf B781} (2018) 568,
{\tt arXiv:1803.08421\,[hep-th]}.

\bibitem{BuchKrP}
I.L.\,Buchbinder, V.A.\,Krykhtin, A.\,Pashnev,
{\it BRST approach to Lagrangian construction for fermionic massless higher spin fields},
Nucl. Phys. {\bf B711} (2005) 367, {\tt arXiv:hep-th/0410215}.

\bibitem{BuchKr}
I.L.\,Buchbinder, V.A.\,Krykhtin,
{\it Gauge invariant Lagrangian construction for massive bosonic higher spin fields in D dimensions},
Nucl. Phys. {\bf B727} (2005) 537, {\tt arXiv:hep-th/0505092}.


\bibitem{BuchKrTak}
I.L.\,Buchbinder, V.A.\,Krykhtin, H.\,Takata, {\it BRST approach
to Lagrangian construction for bosonic continuous spin field}, {\tt
arXiv:1806.01640\,[hep-th]}.



\end{thebibliography}
\end{document}